# The use of the fractal Brouers-Sotolongo formalism to analyze the kinetics of drug release

F. Brouers[a], Tariq J. Al-Musawi[b,*]

[a] Faculty of Applied Sciences, Liege University, Belgium, E-mail address: fbrouers@ulg.ac.be.

[b] Department of Civil Engineering, Faculty of Engineering, Isra University, Amman, Jordan.

* Corresponding author: E-mail address: tariqjwad@yahoo.com, tariq.almusawi@iu.edu.jo. Tel.: 00962 791589708.

**Abstract**

We have applied the Brouers-Sotolongo fractal kinetic equation ($BSf(t,n,\alpha)$), improving notably the precision, to nine cases reported recently in the literature on drug release. The reason of using this equation is that it contains as approximations some of the mostly used empirical formula used in that field. Moreover, this equation is now successfully employed for the investigation of sorption of contaminants in aqueous media. An important extension of the $BSf(t,n,\alpha)$ has been the introduction of variation of the fractal time coefficient ($\alpha(t^\nu)$). This improvement can lead to a greater precision of the fits and deduce some hint on the nature of the drug release process which can give precious information to propose microscopic molecular ad hoc models. We, therefore, suggest the use of the $BSf(t,n,\alpha(t^\nu))$ formula, as a first step, in any detailed investigation and practical application of drug release data both in vitro and in vivo studies starting with the Weibull and Hill approximations to follow properly the physical solution.

# 1. Introduction

Drugs are a big part of the pharmaceutical compounds with extensive application in the treatment of human, fish, and poultry infections (**Kalhori et al., 2017; Soori et al., 2016**). In case of humans, the vast majority of drugs are manufactured in the form of tablets, capsules or suspensions which are taken orally (**Al-Musawi et al., 2019; Swarbick and Boylan, 1996**). At present, the manufacturing high-quality drugs has become a top priority of the pharmaceutical companies, therefore, several studies has been conducted by researchers in order to best describe of drug release process in body. Drug release is an important concept in the pharmaceutical sciences, it is defined as the mechanism of a drug molecule leaves its dosage form, and then it is subjected the processes of dissolution (desorption from surface), distribution in the release medium, metabolism and eventually becoming available for target sites of pharmacological action within the body (**Singhvi and Singh, 2011**).

The study on the mathematical description of the absorption kinetics of drug release is a basic problem in pharmacokinetics and of importance to compare the therapeutic performance of different drugs, for example, generic and innovative ones (**Sopasakis et al., 2018; Fernandez et al., 2011; Raval et al., 2010; Pereira, 2010**). From another side, there is a necessity for the experimental data of drug release to be modeled and fitted with the used mathematical models in order to successfully analyze and optimize the design of the drug delivery systems. For many cases, the drug dissolution is the most important and predominated step in drug release in the body (**Siepmann and Siepmann, 2012**). Until recently, to study the drug dissolution profile in the form of its cumulative absorption i.e. the direct comparison of the experimental curves and the use of mathematical models to compare their statistical parameters, most authors have used well-known empirical equations derived from the theory of chemical kinetic reactions (linear, linearized, or nonlinear equations) (**Selmi et al., 2018a; Sopasakis et al., 2018; Rostamizadeh et al., 2018**). As in many similar problems, the two most popular equations in that field are the Weibull (**Kosmidis and Macheras, 2018; Barzegar-Jalali et al., 2008; Papaadopoulou et al., 2006**) and the Hill (**Goutelle et al., 2006; Korhonen et al., 2004; Mahayni et al., 2000**) equations. Other equations have been used in the literatures which are: zero, first, and second order, Higuchi, Hixson-Crowell, Korsmeyer-Peppas, Michaelis-Menten, and reciprocal powered time model equations (**Singhvi and Singh, 2011; Barzegar-Jalali et al., 2008; Ramteke et al., 2014**). We have cited some of the above mentioned equations for reasons which will appear later in the subsequent sections of the present paper.

Other authors have developed more sophisticated (microscopic or mesoscopic) model based on the classical theory of diffusion to analyze drug release when microporous or mesoporous surfaces were involved (**Vo et al., 2018; Barzegar-Jalali et al., 2008**). These models are based on the consideration of the various diffusion steps, the chemical or physical nature of the interaction at the liquid-solid interface, the structure of the drug molecule, and the surface characteristics of the porous media. Generally, when used to fit experimental data, these models end up with approximations which are equivalent to some of the empirical formulas quoted above. One of the reasons is that there is no one to one correspondence between the more detailed and the empirical models which finally have to be used. The Weibull equation, the best example being, that can approximate and represent a number of different physical situations. Therefore, finding a methodology for modeling and fitting with precision the drug release curves which can unify and improve the above mentioned models is of great interest. In recent years, the application of fractal models like Brouers-Sotolongo kinetics formula (**Brouers and Sotolongo-Costa, 2006; Brouers, 2014**) in the modeling the kinetic diffusion data of organic and inorganic contaminants in the solid-liquid phases has been very satisfactory (**Selmi et al., 2018a; Selmi et al., 2018b; Brouers and Al-Musawi, 2018**). Indeed, Brouers-Sotolongo model is based on a Burr-XII rigorous statistical function (**Burr, 1942; Singh and Madala, 1976**) (a function equivalent to the deformed exponential introduced by Tsallis in its theory of non-extensive entropy (**Tsallis, 2002; Brouers et al., 2004; Picoli et al., 2003; Brouers and Sotolongo-Costa, 2005**)). It is worth mentioning that the use of the Burr-XII distribution in chemical reactions in complex systems has been given theoretical ground in the works of the Wroclaw school based on stochastic and statistical theories (**Jurlewicz and Weron, 1999**).

Due to the evolution in the computer software, many programs have been developed by scientific workers to solve, for example, the engineering problems. Recently, specific computational softwares have also been introduced for fitting process, particularly for kinetic dissolution curves. One of the more recent is the "kinetDS" one software proposed by (**Mendyk et al., 2012; Krupa et al., 2018**). This software was primarily designed for handling pharmaceutical dissolution tests. In addition, the nonlinear fitting algorithms built in the Mathematica technical computing software, in which the full calculation of the present study is done, is also widely utilized in the field of the modeling of the experimental data.

The purpose of this paper is to show that the Brouers-Sotolongo model contains in itself all the mentioned empirical functions and get better precision than most of the methods used so far in field of drug release. We will thus compare the obtained results of our theory with those of a few recent papers which have employed empirical, semi-empirical, and computational methods, that some of them using a heavy mathematical formulation. We will show that in each of these cases, if we introduce a time variation of the fractal coefficient, we get better precision and reach quite similar physical conclusions.

## 2. Theoretical model

The Brouers-Sotolongo model has been used successfully, for example, quoting some of the most recent papers (**Al-Musawi et al., 2017; Figaro et al., 2009; Unuabonah et al., 2019; Wakkel et al., 2019; Selmi et al., 2018a; Mbarki et al., 2018; Selmi et al., 2018b**) in the field of sorption in aqueous phase on porous, micro and macro materials. Theoretically, Brouers-Sotolongo model has been legitimized by stochastic arguments (**Jurlewicz and Weron, 1999**). The Brouers-Sotolongo model, moreover, can provide simple expressions for the half-life sorption or release time scale as well as for parameter of the time-dependent rate which can give hints for the changes of the physico-chemical processes that appear during the evolution of the dissolution process. To demonstrate the usefulness of Brouers-Sotolongo model, we have analyzed the dissolution data of drug release that discussed in the original paper of (**Mendyk et al., 2012; Krupa et al., 2018**) and some representative published data (**Vo et al., 2018; Siepmann and Peppas, 2001; Schlupp et al., 2011; Rostamizadeh et al., 2018**).

In the present study, the proposed Brouers-Sotolongo kinetic formula has been named BSf(t,n,α). The generalized form of BSf(t,n,α) model is a solution of the following equivalent fractal differential equations (eqs. 1 or 2):

$$\frac{dq(t)}{dt^a} = -\frac{1}{\tau} q(t)^n \qquad (1)$$

$$\frac{dq(t)}{dt} = -\frac{a t^{a-1}}{\tau} q(t)^n \qquad (2)$$

Where q(t) is the quantity of drug released or dissolved molecules at specific sorption or release time (t, hr or day), τ is a characteristic time (hr or day), n is a parameter denoting to the fractional order of the reaction, and α is an essential constant representing the fractal time coefficient and can be used to macroscopically express the system complexity.

In addition, BSf(t,n,α) model is a general form of the well-known first order differential model (eq.3), which gives the exponential function.

$$\frac{dq(t)}{dt} = -\frac{1}{\tau} q(t) \qquad (3)$$

The solution of eq.1 depends on the limiting conditions, as shown below:

If $q(0) = q_i$ and $q(\infty) = q_f$ in which $q_f \ll q_i$ or $= 0$,

Where $q_i$ is the initial drug quantity, and $q_f$ is the quantity of drug which has not been released at the end of the process, one has the decay solution as shown in eq.(4), as follows:

$$BSf(t, n, \alpha) \equiv q(t) = (q_i - q_f)\left(1 + (n-1)\left(\frac{t}{\tau}\right)^\alpha\right)^{-\left(\frac{1}{n-1}\right)} + q_f \qquad (4)$$

If $q(0) = 0$ and $q(\infty) = q_m$

where $q_m$ being the maximum released quantity of drug, one has the cumulative solution (eq.5):

$$\frac{q(t)}{q_m}(\%) = 1 - S(t) \qquad (5)$$

With $S(t) = \left[1 + (n-1)\left(\frac{t}{\tau}\right)^a\right]^{-\frac{1}{n-1}} \qquad (6)$

Where $S(t)$ has the form of the survival part of the cumulative Burr XII distribution function, and can be expresses as the deformed exponential distribution used in the statistical physics of complex systems in particular in the works on the non-extensive entropy (**Tsallis, 2002**).

$$\exp_n[x] = (1 - (n-1)x)^{-1/(n-1)}, \text{ with } \exp_1[x] = \exp[x] \qquad (7)$$

The two asymptotic trends of the $S(t)$ are:

i) For $t \ll \tau$ $\quad S(t) \to 1 - (t/\tau)^a \qquad (8)$

ii) For $t \gg \tau$ $\quad S(t) \to (t/\tau)^{-\left(\frac{a}{n-1}\right)} \qquad (9)$

When $\left(\frac{\alpha}{n-1}\right) < 1$, the Burr XII function belongs to the basin of attraction of the heavy tail Lévy distributions.

The fractal differential equations (1 and 2) can be rewritten in the form of a first order differential equation as shown in Eq.10.

$$\frac{dq(t)}{dt} = -R(t)q(t) \qquad (10)$$

Where $R(t)$ is a time dependent rate (or hazard function in reliability theory) (or intensity of transition in relaxation theory) (**Jurlewicz and Weron, 1999**; **Brouers and Sotolongo-Costa, 2006; Brouers, 2014; Kosmidis and Macheras, 2018; Papaadopoulou et al., 2006**), which is calculated by using the following equation:

$$R(t) = -\frac{d}{dt}\ln(S(t)) = \frac{1}{\tau(t)} = \frac{\alpha}{\tau}\frac{(t/\tau)^{\alpha-1}}{(1+(n-1)(\frac{t}{\tau}))} \qquad (11)$$

It is worth to note that the half time $\tau_{50\%}$ (hr or day) corresponding to $q(t) = 0.5\, q_m$ is given by:

$$\tau_{50\%} = \tau\left(\frac{(0.5)^{-n+1}-1}{n-1}\right)^{1/\alpha} \qquad (12)$$

One of the interesting property of the BSf(t,n,α) formula is that it can yield most of the empirical formulas encountered in the literature. These empirical formulas can be expressed as approximations of the full BSf(t,n,α) equation giving them a more precise statistical signature, these formulas are: BSf(t,0,1) is zero- order kinetics equation; BSf(t,1,1) is-first order kinetics equation; BSf(t,2,1) is second order kinetics equation; BSf(t,1,α) is the Weibull kinetics equation; BSf(t,2,α) is the Hill (or log-logistic) kinetics equation; BSf(t,2,1) is the Michaelis-Menten equation; BSf(t<<τ,1,α) is the Korsmeyer-Peppas kinetics equation, and BSf(t<<τ,1,1/2) is the Higuchi kinetics equation. Notably, a lag time can be introduced in these formulas if it is necessary.

In some of our previous papers on sorption in porous materials, we had to introduce a time dependent fractal coefficient to account for the change of sorption conditions with time. This improvement of the theory can be used in this context too in order to achieve a better fit of the release curves. We assumed a power law behavior which is common in complex systems in the following form:

$$\alpha(t) = \alpha_0 + (\alpha_s - \alpha_0(t/t_s)^\nu) \qquad (13)$$

Where $\nu$ is an empirical parameter, $t_s$ a time corresponding to the experimental saturation of the release, $\alpha_0$ and $\alpha_s$ are the initial and at saturation time values of the power law, respectively.

With this expression the solution of eq.(11)

$$R(t) = \left(\frac{t}{\tau}\right)^{\alpha(t)} \left(\frac{1}{t}a(t) + \frac{1}{t_s}(a_s - a_0)\left(\frac{t}{t_s}\right)^{\nu-1} \ln\left(\frac{t}{t_s}\right) / (1 + (n-1)\left(\frac{t}{t_s}\right)^{a(t)}\right) \qquad (14)$$

With $a(t)$ given by eq.(13), in addition, Eq.(14) reduces to eq.(11) when $a_s = a_0$.

In the following, when we introduce a time dependent fractal coefficient (eq.13), $BSf(t,n,\alpha)$ will be written $BSf(t,n,\alpha(t^\nu))$. When n=1 or n= 2, we will call it Weibull or Hill, respectively. The solver software of $BSf(t,n,\alpha(t^\nu))$ has been built with the nonlinear fit of the program Mathematica (version 10). It is worth to mention that the accuracy of the fitting between the data and used models was evaluated in the light of the determined values of the regression coefficient, which were presented in six decimal places.

## 3. Results and discussion

### 3.1 Analysis the data of Cases 1-3

**Mendyk et al., (2012)** have proposed an algorithm called "KinetDS" to analyze drug dissolution test data. They have applied it to three different cases to assess its usefulness. Alternatively, we have applied the $BSf(t,n,\alpha(t^\nu))$ formalism to treat these three cases and obtained a clearly better fit and a better characterization of the relevant parameters. The first is the analysis of a simulation of data obtained from an exact Hill equation and reproduced in Table 4 of their paper (case 1). The second is the data obtained from the application of the KinetDS software to the simulation of a biphasic drug release consisting in a first rapid phase followed by a second exhibiting a linear evolution (case 2), these data are listed in Table 5 in **Mendyk et al., (2012)**. The third is a real case situation, it is to say, the results published by **Patel et al., (2008)** in dissolution technologies (case 3) which were listed in Table 6 in **Mendyk** et al., (2012). For this purpose the nonlinear modeling methods of Mathematica program was utilized for the determination of the model parameters. The analysis for the dissolution data of case 1 demonstrated that our methodology was reproduced the parameters of Hill model as accurately ($R^2=1$) than those obtained from **Mendyk et al., (2012)** analysis (Table 1). Indeed, the data of dissolution profile (case 1) were presented perfectly by using Hill model as depicted in Figure 1.

Table 1 listed the results of analysis of the two parts of case 2 data. At first, the data of each part were fitted separately and the results of fitting are depicted in Figure 2a. Secondly, we have fitted the whole curve (Figure 2b). All the models fit the data with $R^2$ values> 0.999 (Table 1). It can be noticed that Weibull model was the best fitted model for each part of case 2 data, while, the whole data of these two parts fitted well with $BSf(t,n,\alpha(t^\nu))$ model.

The results of fitting of case 3 by **Mendyk et al., (2012)** showed that the Weibull model was satisfactory represented the data with $R^2$ values of 0.9496 and 0.9541, respectively. The analysis of this study for case 3 data (Table 1) demonstrated that it can be reached to $R^2$ values higher than 0.999. Moreover, $BSf(t,n,\alpha(t^\nu))$ model exhibited high precision values compared with the other used models. The data of case 3 and the best fitted curve is being shown in Fig. 3a, as well as, the rate results are depicted in Fig. 3b.

Table 1. The modeling results of cases 1, 2, and 3 drug release data

| **Model** | $q_m$ | n | $\alpha(t^\nu)$ | $\upsilon$ | $\tau_{50\%}$ | $R^2$ |
| --- | --- | --- | --- | --- | --- | --- |
| Case 1 | | | | | | |
| Weibull | | 1 | 3.37 | | 6.47 | 0.99837 |
| Hill | | 2 | 5 | | 6.31 | 1 |
| Case 2 | | | | | | |
| Weibull (First part of Fig. 2a) | | 1 | 0.30 | | | 0.999991 |

| | | | | | |
|---|---|---|---|---|---|
| Weibull (second part of Fig. 2a) | | 1 | 0.71 | | 0.999960 |
| Hill | | 2 | 0.68 | 1.67 | 0.998744 |
| Weibull | | 1 | 0.45 | 1.87 | 0.999769 |
| BSf(t,n,α(t$^ν$)) | | 1.6 | 0.3-0.71 | 0.2 | 0.999911 |
| Case 3 | | | | | |
| Weibull | 100 | 1 | 0.91 | 4.07 | 0.999906 |
| Hill | 100 | 2 | 1.66 | 4.09 | 0.999221 |
| BSf(t,n,α(t$^ν$)) | 100 | 1.15 | 0.61-1.24 | 0.5 | 0.999990 |

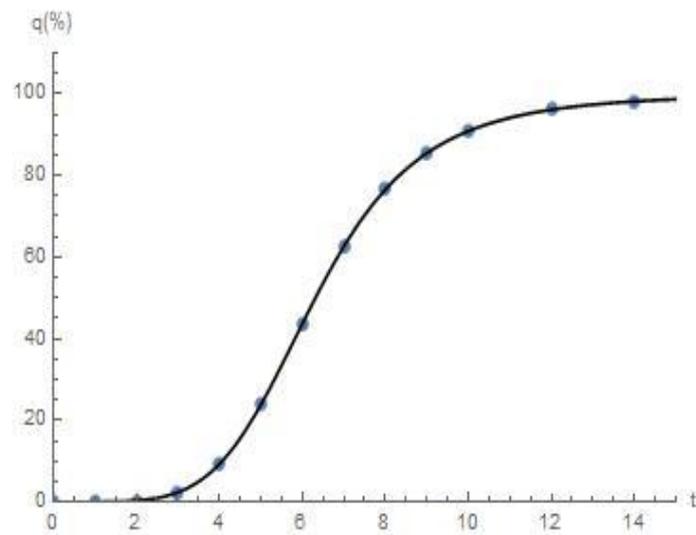

Figure 1. Fit of case 1 data (dotted points) with the Brouers-Sotolongo model (solid line)

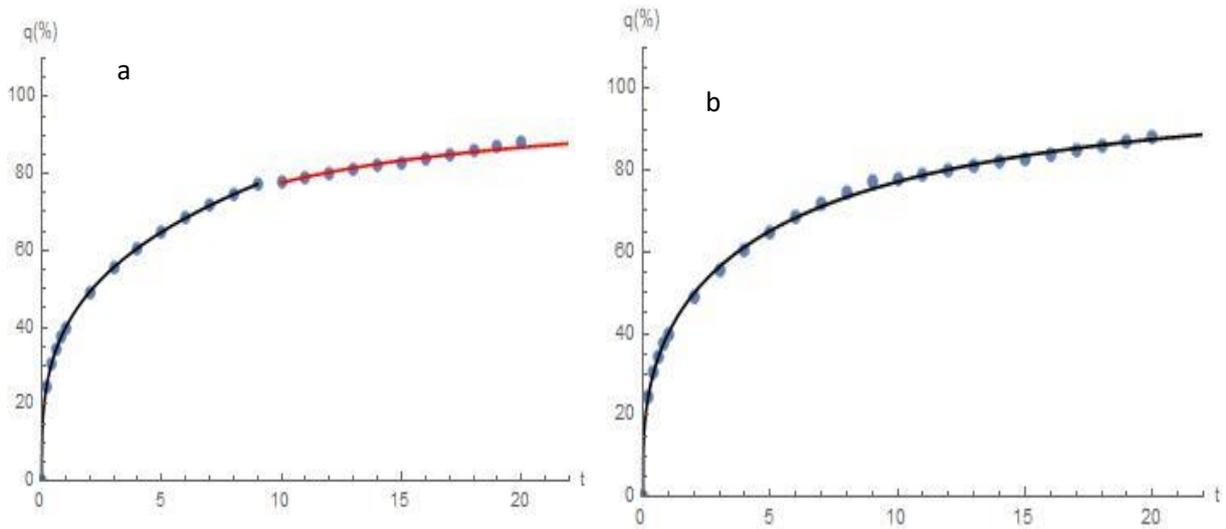

**Figure. 2.** Fit of separately the two parts of the case 2 data (dotted points) with the Brouers-Sotolongo model (solid line) (a), and fit the whole case 2 data with the same model (b)

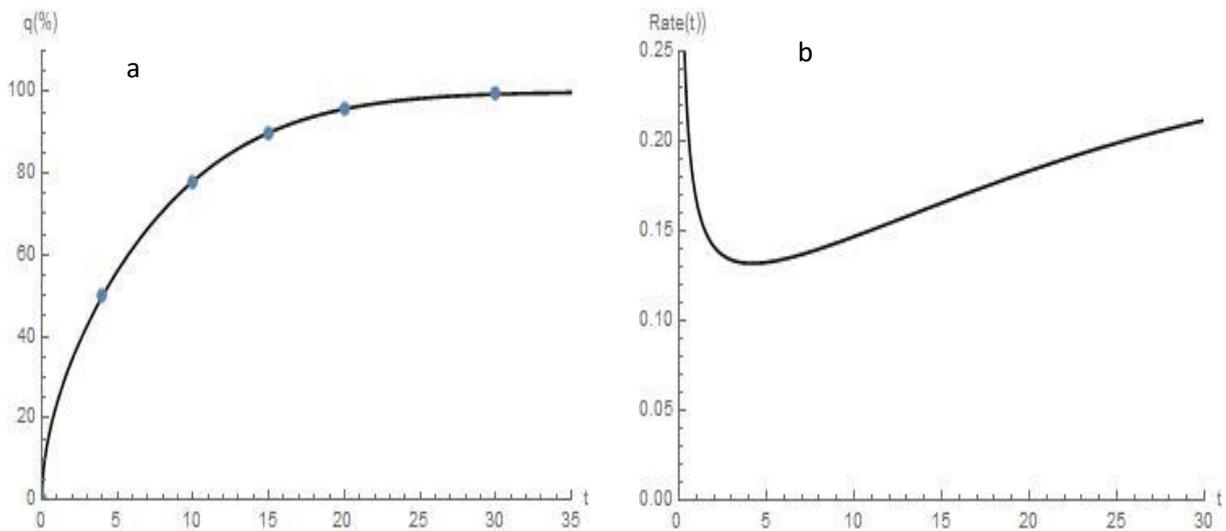

**Figure 3.** Fit of case 3 data (dotted points) with the Brouers-Sotolongo model (solid line) (a), and the rate results (b)

## 3.2 Analysis the data of Cases 4-6

In cases 4 and 5, we have applied the $BSf(t,n,\alpha(t^\nu))$ method to the experimental data of the upper curves of Fig.6a (case 4) and Fig.6b (case 5), respectively, of reference (**Siepmann and Peppas, 2001**). These authors have discussed the application of a set of usual empirical and semi-empirical formulas to model drug release from delivery systems based on hydroxypropyl methylcellulose. They did not give any values of $R^2$, but the comparison of their and our fitting shows clearly that the $BSf(t,n,\alpha(t^v))$ method is with values larger than 0.9999 yields a better representation of the evolution of the kinetics of the release and provides relevant and useful parameters as tabulated in Table 2. In addition, it can be abvioused from Figures 4 and 5 which depict the evolution curves corresponding to $BSf(t,n,\alpha(t^v))$ fitting with the the real data of cases 4 and 5, respectively, that these two types of data can be fitted precicisally from inception to saturation condition by our theory. Also, the results indicated that the improvement in $BSf(t,n,\alpha(t^v))$ model by introducing the time dependent variable ($t_s$) was necessary in studied cases to achieve a good fit of the drug release data.

**Vo et al., (2018)** have used a complex mathematical model based on diffusion mechanisms to analyze the drug release from polymer-free coronar stemts with micorporous surfaces (case 6). This model predicts two stages of the release profile, which are a relatively initial rapid release of almost the entire drug followed by a slower process of the release of the remaining drug quantity. The first stage, in fact, is a diffusion process, while the second slow stage is due to an adsorption-desorption mechanism between the drug molecules in the release medium and the solid drug surface. The adsorption-desorption process occurs close to drug surface region depending on its surface roughness, which is can significantly affected the drug release amount via the effects of Van der Waals forces (see Fig. 2 in **Vo et al., (2018)**). Referring that the manufactured drug whose surface is rough has a large surface area, which is an important factor that can retard or improve the desorption rate of drug molecules from its exposure surfaces to desorption medium (**Gultepe et al., 2010; Sulaymon et al., 2013**). The model devised by **Vo et al., (2018)** and used to fit the experimental data presented in their paper has two drug release stages. The first is a rapid initial classical diffusive regime model until a characteristic time defining the transition t (this release was continued for a period of about 0.5 day). The second stage, an adsorption-desorption regime influenced by the surface, is described by a sophisticated model depending on microscopic and mesoscopic geometrical and physical parameters of the system and results from the tedious solution of a set of differential and nonlinear equations. The second stage of the remaining drug release is relatively slow and extended for a period of about a week.

We have analyzed these data with the $BSf(t,n,\alpha(t^\upsilon))$ method and the results showed that we can achieve as in the previous casas a precision $R^2 > 0.999$ (table 2), moreover, it can be observed from Fig.6a that $BSf(t,n,\alpha(t^\upsilon))$ captured the case 6 drug release data well. All the determined values of α is less than 1, suggesting fractal release process of the drug (**Selmi et al., 2018b**). Also, the calculated rate results according to eq.11 are displayed in Fig. 6b. The calculation of the time-dependent rate exhibits as in the **Vo et al., (2018)** paper a rapid followed by a much slower release rate. The curve exhibits a slight "bath tub" behavior due to the change of release mechanism, surface to volume sorption.

Table 2. Results of the modeling of cases 4, 5, and 6 data

| Model | n | α | υ | $t_s$ | $\tau_{50\%}$ | $R^2$ |
|---|---|---|---|---|---|---|
| Case 4 | | | | | | |
| Weibull | 1 | 1.27 | | | 3.28 | 0.998517 |
| Hill | 2 | 2.03 | | | 3.12 | 0.994740 |
| $BSf(t,n,\alpha(t^\upsilon))$ | 1.08 | 0.88-2.20 | 1.2 | 11 | | 0.999979 |
| Case 5 | | | | | | |
| Weibull | 1 | 1.76 | | | 4.53 | 0.998898 |
| Hill | 2 | 2.7 | | | 4.35 | 0.999185 |
| $BSf(t,n,\alpha(t^\upsilon))$ | 1 | 1.4-2.7 | 2 | 11 | | 0.999962 |
| Case 6 | | | | | | |

| | | | | | |
|---|---|---|---|---|---|
| Weibull | 1 | 0.37 | | 0.040 | 0.998339 |
| Hill | 2 | 0.61 | | 0.037 | 0.999058 |
| BSf(t,n,α(t$^v$)) | 1.95 | 0.57-0.74 | 2 | 10 | 0.999271 |

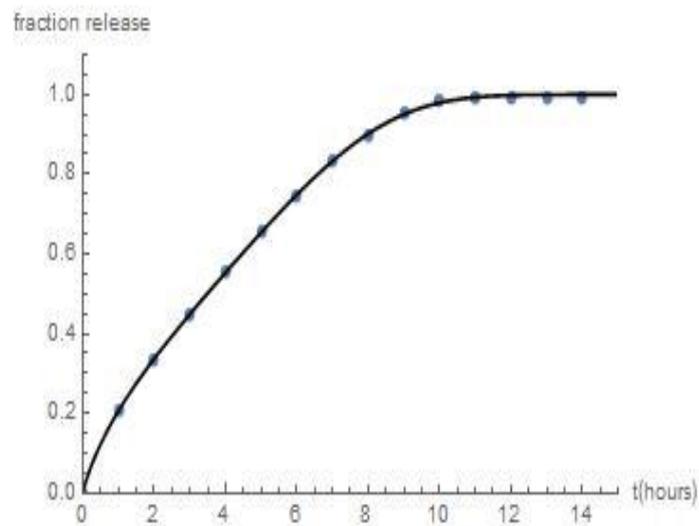

Figure 4. Fit of case 4 data (dotted points) with the Brouers-Sotolongo model (solid line)

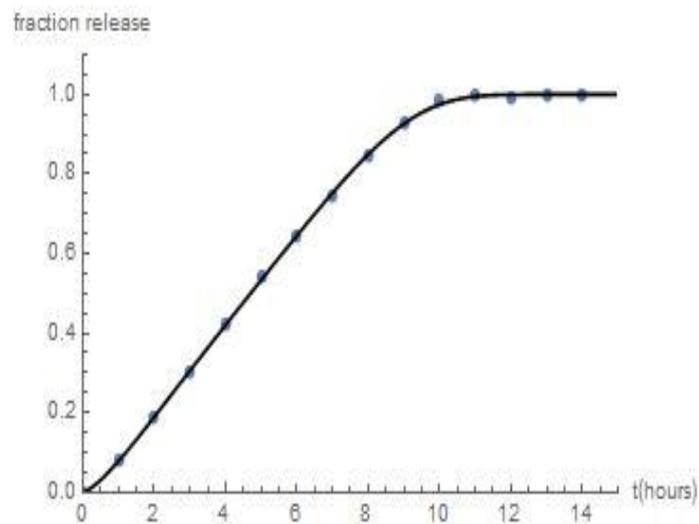

Figure 5. Fit of case 5 data (dotted points) with the Brouers-Sotolongo model (solid line)

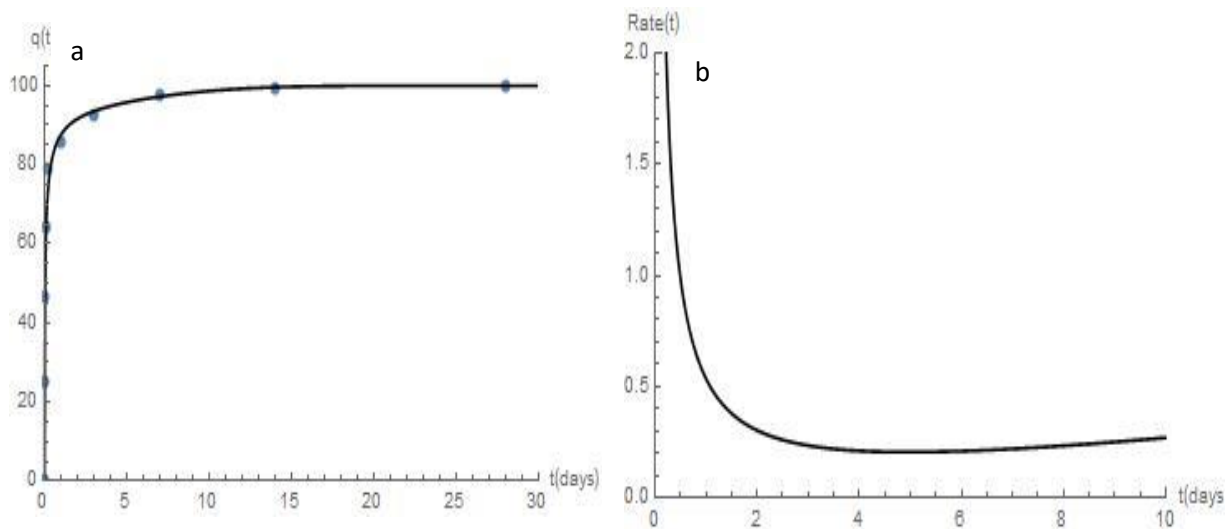

Figure 6. Fit of case 6 data (dotted points) with the Brouers-Sotolongo model (solid line) (a), and the rate results (b)

**3.3 Analysis the data of Cases 7-9**

In case 7, we have analyzed Glucocorticoid drug release data from solid lipid nanoparticles with the time of release up to 8 hr, in particular the Figure 2 of study conducted by **Schlupp et al., (2011)**. The authors concluded that these data are described by a two successive linear fit. The first line started from initial to 3 hr release time while the second line started from 4 to 8 hr release time, with $R^2$ values of 0.980 and 0.983, respectively. We have used the statistical $BSf(t,n,\alpha(t^\nu))$ equation to fit these data and reached similar above conclusions. The obtained value of $R^2$ is much higher than the two values reported by **Schlupp et al., (2011)** (Table 3). In addition, the experimental points are well fitted by the nonlinear curve fitting results obtained from applying $BSf(t,n,\alpha(t^\nu))$ model as depicted in Fig. 7a. The complexity of the system is being represented by a fractal coefficient which is time dependent to represent a gradual change of release regime which is confirmed by the behavior of the rate (Fig. 7b) resulting by a change with time of the fractal parameter which varies from 2.43 to 1.90. The obtained value of $R^2$ is much higher than the two values reported by **Schlupp et al., (2011)** (Table 3).

The last two cases concern the drug delivery of a conjugate of an antitumor conjugated with a copolymer and analyzed recently with Hill and Weibull formula by **Rostamizadeh et al., (2018)**. The authors fitted the experimental data of drug release with six different models and the highest $R^2$ value obtained did not exceed 0.86. In the two cases corresponding to two different media differing from their pH (case 8: the inner release profile of Fig. 8 at pH=7.4; case 9: the inner release profile of Fig.8 at pH=5.5, of that paper), our results are clearly better than those published in that paper (Figs. 8 and 9, and Tables 3). It is clear that the determined value of the statistical accuracy parameter was greater than 0.99 in many cases indicating the efficacy of our method to fit the drug release data.

Table 3. Results of the modeling of case 7, 8, and 9 data.

| Model | $q_m$ | n | α | υ | $t_s$ | $\tau_{50\%}$ | $R^2$ |
|---|---|---|---|---|---|---|---|
| Case 7 | | | | | | | |
| Weibull | 90.44 | 1 | 1.64 | | | 2.17 | 0.999431 |
| Hill | 97.04 | 2 | 2.24 | | | 2.42 | 0.999757 |
| Hill with α($t^υ$) | 100.44 | 2 | 2.43-1.90 | 2 | 10 | | 0.999762 |
| Case 8 | | | | | | | |
| Weibull | 63 | 1 | 0.47 | | | 11.40 | 0.986907 |
| Hill | 63 | 2 | 0.61 | | | 14.90 | 0.984134 |
| Hill with α($t^υ$) | 64 | 1.5 | 0.3-1.38 | 1.5 | 120 | | 0.998129 |
| Case 9 | | | | | | | |
| Weibull | 10 | 1 | 0.77 | | | 10.90 | 0.992279 |
| Hill | 10 | 2 | 1.24 | | | 14.60 | 0.989892 |

| Hill with α(tᵛ) | 11.1 | 2 | 0.90-1.66 | 2 | 100 | 0.999399 |

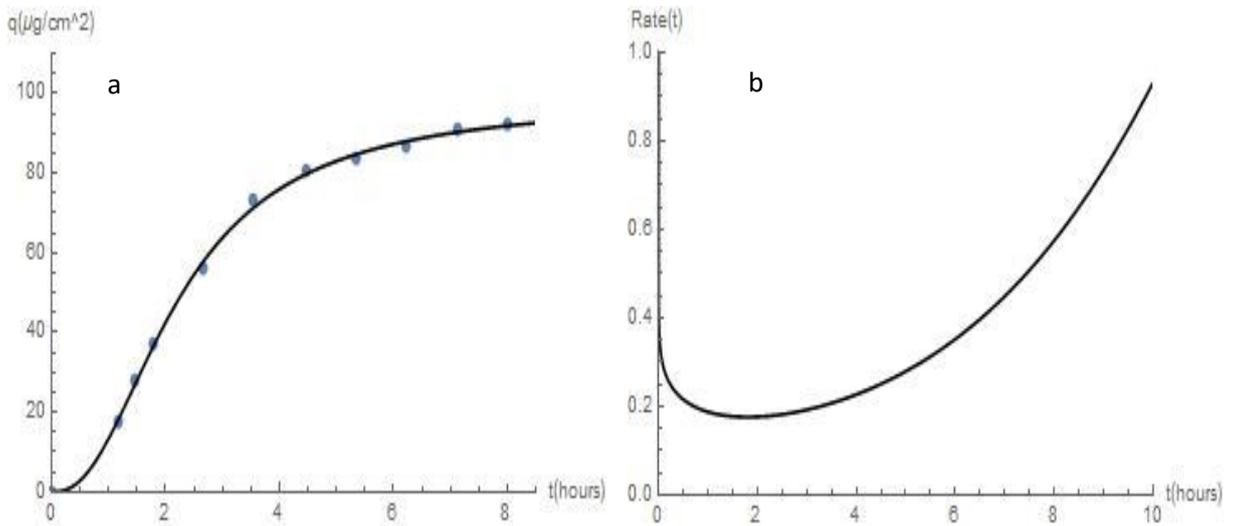

Figure 7. Fit of case 7 data (dotted points) with the Brouers-Sotolongo model (solid line) (a), and the rate results (b)

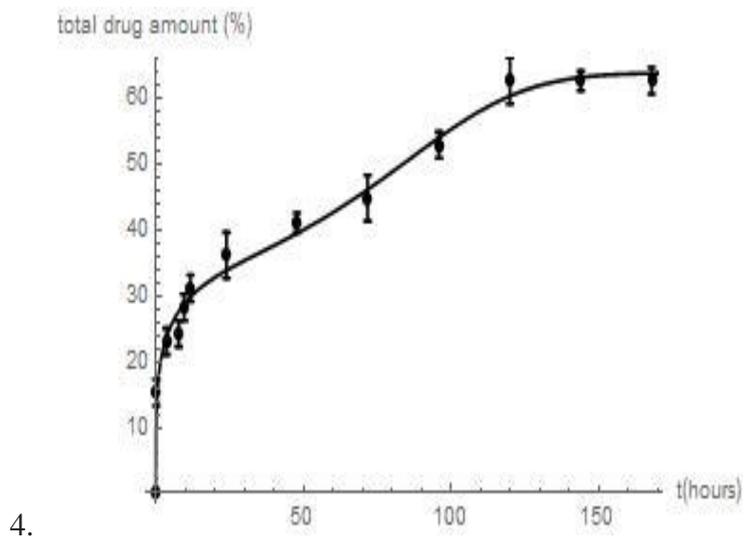

4.

Figure 8. Fit of case 8 data (dotted points) with the Brouers-Sotolongo model (solid line)

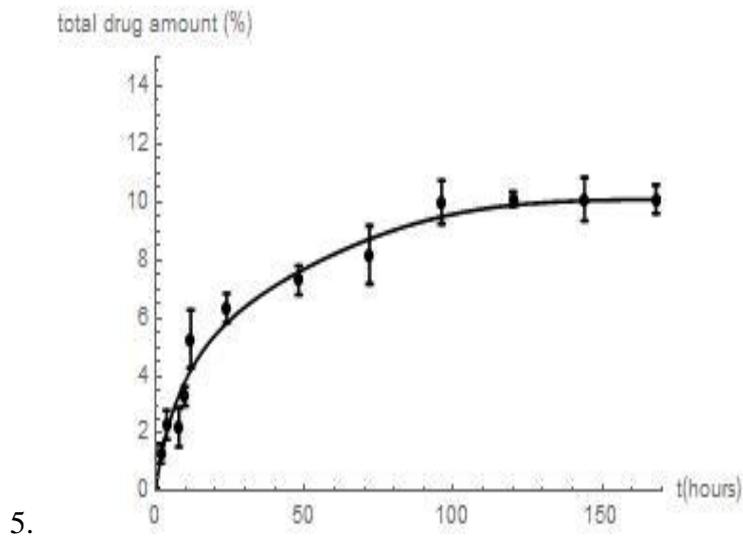

5.

Figure 9. Fit of case 9 data (dotted points) with the Brouers-Sotolongo model (solid line)

**Conclusions**

We have applied the Brouers-Sotolongo fractal kinetics (BSf(t,n,α(t$^v$))) equation to nine cases reported recently in the literature on drug release where were employed traditional semi-empirical formulas and complex software. This formula, when the right approximations are performed, yields some of the commonly uses empirical ones used in this field. The (BSf(t,n,α(t$^v$))) equation can be obtained by fundamental statistical and stochastic arguments. In each case we have obtained a better precision with fits characterized by a squared correlation factor ($R^2$) values between 0.994 and 0.99999 most of the time much higher than the one published in the studied papers. This can be compared also with the compilation of drug release data of (**Barzegar-Jalali et al., 2008**) who reported the results of 106 studies with traditional methods and obtained 15 % of the $R^2$ values higher than 0.994 and 5% with $R^2$ higher than 0.999.

One of the feature of the (BSf(t,n,α(t$^v$))) equation is that it includes a time-dependent fractal coefficient α(t$^v$) which yields a much precise description of the time dependent rate in this problem in particular the values of the half-time release time and the maximum release and allows guesses on the change with time of the physico-chemical nature of the release process. Another important consequence of introducing the time dependence of the fractal coefficient is to obtain a correct description of the time dependence of the rate function. In this paper we have encountered four different situations:

$\alpha_0 < 1$ and $\alpha_s < 1$ (case 6)

$\alpha_0 < 1$ and $\alpha_s > 1$ (cases 3; 4; 8, and 9)

$\alpha_0 > 1$ and $\alpha_s > 1$ (case 5)

$\alpha_0 > \alpha_s$ (case 7)

The behavior of the rate is very different in each case. In the second one, the time variation has a non-monotonous "bath tub" shape which is not possible to obtain with a constant value of the fractal coefficient. The analysis assuming a variation with time of this coefficient allows us to take account of a possible change of nature of the release as a function of time.

Our method can be confirmed, if necessary, with more sophisticated microscopic model keeping in mind that different microscopic models can yield, when averaged, to similar macroscopic ones. We, therefore, suggest the use of the (BSf(t,n,$\alpha(t^\nu)$)) formula, as a first step, in any detailed investigation of release data both in vitro- and vivo studies starting with the Weibull and Hill approximations to be able to follow the physical solution in this nonlinear equation.

## 5. Acknowledgments

Authors express their gratitude to colleagues at Isra University (Amman, Jordan) for the support extended for this work.

Vo T.T., Morgan S., McCormick C., McGinty S., McKee S., Meere M., Modelling drug release from polymer-free coronary stents with microporous surfaces, International Journal of Pharmaceutics, 2018; 544:392–401.

Wakkel M., Khiari B., Zagrouba F., Textile wastewater treatment by agro-industrial waste: Equilibrium modelling, thermodynamics and mass transfer mechanisms of cationic dyes adsorption onto low-cost lignocellulosic adsorbent, Journal of the Taiwan Institute of Chemical Engineers, 2019, article in press. https://doi.org/10.1016/j.jtice.2018.12.014.